\documentclass{article}
\usepackage[utf8]{inputenc}
\usepackage{authblk}
\usepackage{graphicx}
\usepackage{amsmath}
\usepackage{amssymb}
\usepackage{verbatim}
\usepackage{amsthm,amsfonts,bm}
\usepackage{MnSymbol}
\usepackage{cite}
\usepackage{graphicx}
\usepackage{enumerate}
\usepackage{enumitem}
\usepackage{color}
\usepackage[colorlinks=true, allcolors=blue]{hyperref}
\usepackage{dsfont}
\usepackage{pgf,tikz}
\usetikzlibrary{arrows}
\usepackage{rotfloat}
\usepackage{float}
\usepackage{mathrsfs}
\usepackage{soul}

\definecolor{qqqqff}{rgb}{0,0,1}
\definecolor{Red}{rgb}{0.9,0,0}
\definecolor{Blue}{rgb}{0,0,0.9}

\title{Is time the real line?}

\author[1]{B. F. Rizzuti}
\author[2]{L. M. Gaio}
\author[3]{Lucas T. Cardoso}
\affil[1]{Departamento de F\'isica, Universidade Federal de Juiz de Fora, MG, Brazil}
\affil[2]{Programa de Pós-Graduação em Matemática,  Universidade Federal de Juiz de Fora, MG, Brazil}
\affil[3]{Coordenadoria Acadêmica, Universidade Federal de Santa Maria, Cachoeira do Sul, RS, Brazil}
\date{}                     
\newtheorem{definition}{Definition}
\newtheorem{theorem}{Theorem}

\newtheorem{proposition}[theorem]{Proposition}
\newtheorem{cor}{Corolary}

\begin{document}

\maketitle

\section{Introduction}
\label{sec.intro}
\noindent

If we open any textbook on physics, even the basic ones, we can see the graphical representation of time as an oriented arrow. To adopt this point of view means that time is a one-dimensional oriented differentiable manifold. This is a basic premise tacitly accepted by the entire community. To be a manifold, we must ask some technical details about the arrow of time, namely \cite{nakahara},
\begin{enumerate}
	\item Is it a topological space?
	
	\item Is it provided with an atlas $\{ (\mathcal{O}_i, \varphi_i) \}$ of open sets $\mathcal{O}_i$ and homeomorphisms $\varphi_i: \mathcal{O}_i \rightarrow \mathcal{A}_i \subset \mathds{R}$ from $\mathcal{O}_i$ onto an open subset $\mathcal{A}_i$ of $\mathds{R}$?
	
	\item For $\mathcal{O}_i \cap \mathcal{O}_j$, is the map $\psi_{ij}:= \varphi_i \circ \varphi^{-1}_j$ from $\varphi_j(\mathcal{A}_i \cap \mathcal{A}_j)$ to $\varphi_i(\mathcal{A}_i \cap \mathcal{A}_j)$ is smooth?
\end{enumerate}
In other words, and more directly, \textit{is the arrow of time homeomorphic to the real line} $\mathds{R}$\textit{?}  

To talk about such foundations, specially when it comes to time, is always a rather delicate issue. We would like to address it here though. Besides the lack of a profound answer to this query, we were motivated also by the seminal paper \cite{sen_1999}, that explains why the Euclidean line is the same as the real one.  

Of course, this question has already been answered (positively) in some way or another throughout history, nonetheless recent findings may point to possible discreteness of time, with attainable experimental observations \cite{christodoulou_possibility_2020}. Our objective here will be to answer it operationally, that is, by basing it in a series of steps which can be done in a laboratory \cite{logic.modern.1958}. For an example on how to follow this philosophy, see \cite{guilherme.rodrigo.rizzuti.2018, luca.daniel.rizzuti.2019, rizzuti.luca.cristhiano.2019}, where we use this method to define one and multidimensional physical quantities and the topology of the physical space, respectively.

To summarise the process of giving an operational definition of physical quantities, we follow a sequence of steps: find the set of objects for which our quantity should be valid (called domain); define an equivalence relation giving us a notion of equality; partition the domain in equivalence classes; define a sum and a scalar multiplication on the quotient set; label each equivalence class with a number such that the sum and the scalar multiplication have their properties preserved. This process details how we can abstract some property of a given object to a set of numbers where it is  not only easier to work with, but we can make use of distinct mathematical structures that allegedly are in an one-to-one correspondence to the quantities under investigation. Of course, having to label our equivalence classes operationally means that our labels will always be at most rational numbers, due to the inherent experimental precision every measure apparatus possesses. Although this does not present an immediate problem, we lose some very important properties, which are usually taken for granted, like differentiation and convergence of Cauchy sequences, and all mathematical operations related to limit processes. By making a leap in abstraction, we assume some of these properties in order to analyse if the operational philosophy can be used past this point.

Although our main concern here is related to the time structure, the path we take will lead us implicitly to the Galilean space-time. According to the recent comment by A. Staruszkiewicz, ``(it) is, up to date, a topic which is not presented the way it deserves'' \cite{staruszkiewicz_probability_2020}. A short description may be seen in \cite{arnold_mathematical_1989} and a thorough geometric picture on the nature of the Galilean spacetime can also be found in \cite{kopczynski_spacetime_1992}. All of these references already consider, from the beginning, a complete background (in a topological sense), where, for example, the very notions of differentiability can be used. Once again we face this \textit{ad hoc} assumption that both space and time have no holes, and this is the main issue we would like to address in this work.

Our approach here will be somewhat different. In the next Section, after the operational definition for time lapse, leveraging it to the status of a physical quantity, we naturally arrive at the concept of space-time, where one of the main topics of discussion becomes the notion of simultaneity, as our own time definition will depend heavily on this. This initial construction lacks some geometrical meaning, like an orientation. We give thus a step further in Sections 3 and 4, where,  by using free particles, we are able to define simultaneity and, consequently, a partial order for events in the space-time, which is not clear at first. Finally, in the next Section, we use this order to define a topology on the space-time and discuss its properties.

To finish our article, in Sections 6 and 7, we discuss how this method can be mirrored to a relativistic context, as this is the natural follow up of this text. Of course, to do this we must reformulate our notion of simultaneity as many of the problems that arise when we consider speeds near $c$ will appear also with our construction. Section 8 is left for conclusions.

\section{\texorpdfstring{$\delta$}{d}-Time as a physical quantity}

The concept of a physical quantity may be constructed according to an operational approach. It was extensively discussed in Refs. \cite{guilherme.rodrigo.rizzuti.2018, luca.daniel.rizzuti.2019}, which cover most known quantities such as spatial distance, mass, vectors, dual vectors and tensors. Even time may be seen as a physical quantity, as we shall see in a moment. Firstly, let us discuss the very unifying concept of a physical quantity to see how time fits into this description. The initial steps consist of providing the set where the corresponding quantity $G$ may be defined. We call it the domain $D_G$ of the quantity $G$. $D_G$ is then divided into equivalence classes by an experimental equivalence relation $\sim \, \subset D_G \times D_G$. In some sense, this step is connected to the natural human impetus of seeking patterns. Each class tells us what are the $G$-equivalent elements, that is, the elements which are equal according to the $G$-classification. Finally, one associates injectively each class to the corresponding set of values $\mathds{V}_G$. The set $\mathds{V}_G$ is equipped with a sum and multiplication by numbers, after all, it is natural to make predictions, comparisons, precise measures, etc in quantitative sciences. These two operations, which satisfy good enough properties, graduate $\mathds{V}_G$ to a vector space. For the particular case where $\mathds{V}_G$ is one-dimensional, the own dimension\footnote{As of Linear Algebra.} of the quantity turns out to be a basis for $\mathds{V}_G$. For instance, any distance between a pair of points may be written as a real number multiplied by the vector $1m$.

Following this prescription, our first step consists of providing the very set where the quantity time is defined. It is done in complete analogy with geometry. In fact, just as the quantity distance is defined by a pair of points, time will be defined for a pair of what we shall call an event. (In this sense, we are not actually defining time, but time interval. We shall refer to it as temporal distance or lapse of time, if we may. This is why we wrote a $\delta$ in the title of the Section). Formally, 
\begin{definition}\label{definition.01}
	An \textbf{event} is an occurrence that takes place in a small region of space approximated by a point and it is so fast that its duration is of an instant. 
\end{definition}

The attentive reader may find themselves in a loop. The words ``fast'' and ``duration'' are clearly related to what we already know about time interval and could not be used beforehand. However, we are pretty much used to events on a daily basis, such as the blink of an eye or a flash of light.
Besides that, we also have utilized some concepts related to geometry, \textit{e. g.}, point and space, which can be properly defined, see \cite{rizzuti.luca.cristhiano.2019}.

The definition of an event suggests a natural name to
\begin{definition}
	The set of all events is called \textbf{space-time} and shall be denoted by $\mathscr{ET}$.
\end{definition}

With that being said, the domain of the quantity $\delta$-time, denoted by $D_{\delta T}$, is given by
\begin{equation}
	D_{\delta T}:= \mathscr{ET} \times \mathscr{ET}.
\end{equation}

\begin{figure}
	\centering
	\includegraphics[scale=0.25]{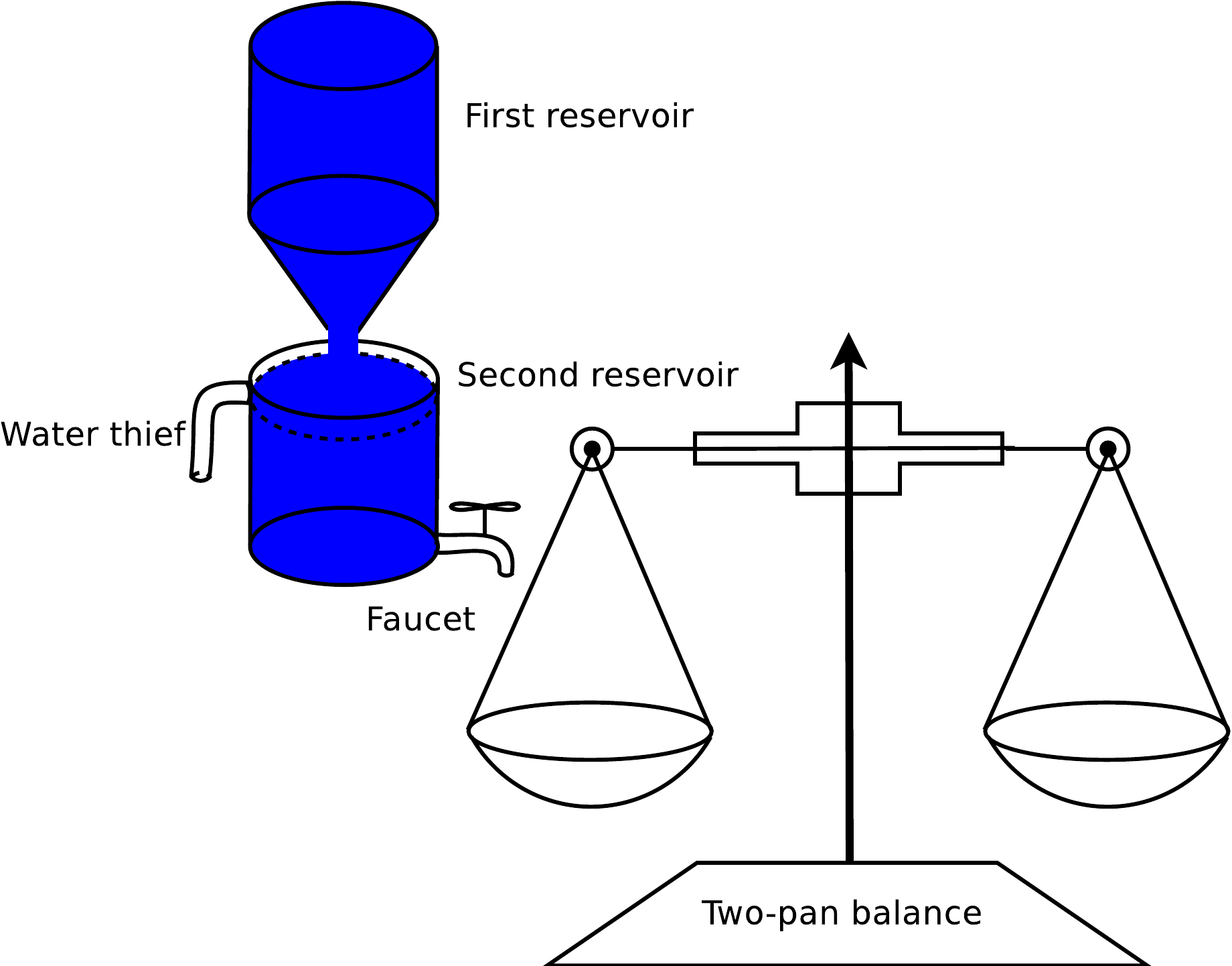}
	\caption{Schematic representation of the clepsydra.}
	\label{Fig_clep}
\end{figure}

The next step in our construction consists of dividing $D_{\delta T}$ into equivalence classes of $\delta T$-equivalent pairs of pairs of events. We are comparing in fact two pairs of events, which explains the repetitive words in the last statement. That is, each class tells us which pairs (of pairs of events) have the same duration. Although somewhat old-fashioned, our method to accomplish this works out well enough. We use here a modern clepsydra \cite{ancient}. The device consists of a water reservoir that feeds a second one, which is kept always with the same amount of water, even when its faucet is open, see Fig. \ref{Fig_clep}. It guarantees regularity in the flow of water.
The second reservoir also  possesses a water thief or drain, responsible for keeping the water level fixed, while the faucet is closed. 
Thus, when the first event $e_1$ happens, one opens the faucet, which is promptly closed as the second event $e_2$ takes place. The faucet spills a mass $m_{12}$ of water that is captured by one of a two-pan scale. Given two pairs of events $(e_1, e_2)$ and $(e_3, e_4)$, we say that they have the same time interval when 
\begin{equation}\label{equiv.rel}
	\delta t(e_1, e_2) \sim_T \delta t(e_3, e_4) \Leftrightarrow m_{12} =_B m_{34}.  
\end{equation}
By $=_B$ on right hand side we mean that both masses $m_{12}$ and $m_{34}$ placed together on the two pans, keep the scale even. Clearly, the process so expressed by Eq. \eqref{equiv.rel} is subjected to the well-definiteness of mass measurability. We will accept it here; see the first comment at the end of this Section. 

\begin{proposition}
	$=_B$ is an equivalence relation. 
\end{proposition}
\noindent
\textit{Proof.} It is a simple experimental task to check that the equality $=_B$ provided by the scale is symmetric, reflexive and transitive. 

\qed

We point out that this proposition, due to the very definition \eqref{equiv.rel}, implies that $\sim_T$ is a equivalence relation as well.

Now we look for a periodic and reproducible pair of events and measure the corresponding mass to fix a standard. For example, $e_1$ may be the beginning of the sunset, when the sun starts to disappear below the horizon, whereas $e_2$ is its end, when the sun is not visible anymore.  We call this corresponding mass $M$
\begin{equation}
	\delta t(e_1, e_2) := 120\,seconds.
\end{equation}

The prescription given above characterizes the time-lapse of all pair of events, say, $e$ and $f$, such that the amount of water accumulated by the pan on the scale when the faucet has been kept open during $e$ and $f$, weighing the mass $m_{12}$. It can be used to numerically characterize all classes of the quotient space  
\begin{equation}\label{equiv.rel.def}
	D_{\delta T}/ \sim_T =  \bigcup_{(e_1,e_2)\in D_{\delta T}} [(e_1,e_2)],
\end{equation}
where $[(e_1,e_2)]:=\{ (e,f)\in D_{\delta T}\vert (e,f) \sim_T (e_1,e_2) \}$.
Actually, we have just defined a map that may be seen as a measure of time intervals
\begin{align}
	\delta t: D_{\delta T}/ \sim_T &\longrightarrow \mathds{V}_{\delta T} \cr 
	[(e_1, e_2)]&\longmapsto \delta t[(e_1, e_2)] = \frac{120s}{M}m_{12},
\end{align}
and the procedure of weighing masses defines the set of values $\mathds{V}_{\delta T}$. It is worth mentioning that this map is well-defined, which can be seen directly from the definition of classes in $D_{\delta T}/ \sim_T$, see \eqref{equiv.rel} and \eqref{equiv.rel.def}.   

Throughout this article, we will use both notations, $\delta t(e_1, e_2)$ for a particular time interval for the pair $(e_1, e_2)$ and $\delta t [(e_1, e_2)]$ for the time interval that characterizes the entire class.    

\begin{proposition}
	The map $\delta t$ is injective. 
\end{proposition}
\noindent
\textit{Proof.} Let $[(e_1, e_2)]$ and $[(e_3, e_4)]$ be two distinct classes in $D_{\delta T}/ \sim_T$, that is, $m_{12} \neq m_{34}$. Thus, $\delta t[(e_1, e_2)] = \frac{120s}{M}m_{12} \neq \frac{120s}{M}m_{34}= \delta t[(e_3, e_4)]. $ 

\qed  

To complete the construction of the physical quantity, we equip $\mathds{V}_{\delta T}$ with a sum and a number multiplication. These two operations are naturally imported from the quantity mass and may be constructed experimentally. 

Let $\delta t(e_1, e_2)$ and $\delta t(e_3, e_4)$ be two time intervals, with corresponding masses $m_{12}$ and $m_{34}$. So, \begin{align}
	+ : \mathds{V}_{\delta T} \times \mathds{V}_{\delta T} &\longrightarrow \mathds{V}_{\delta T} \cr 
	(\delta t [(e_1, e_2)], \delta t [(e_3, e_4)])  &\longmapsto \delta t [(e, f)]
\end{align}
where $[(e,f)]$ is the class corresponding to the inverse image 
$$\delta t^{-1}\left ( \frac{120s}{M}(m_{12}+ m_{34}) \right )$$ 
and $m_{12}+m_{34}$ is obtained by weighing the two amounts of water together. 

Continuing on, the number multiplication is defined as follows. First,
\begin{align}
	\cdot : \mathds{N} \times \mathds{V}_{\delta T} &\longrightarrow \mathds{V}_{\delta T} \cr 
	(n, \delta t [(e_1, e_2)]) &\longmapsto n\cdot \delta t [(e_1, e_2)]:= \delta t [(e_1, e_n)]
\end{align}
and $[(e_1, e_n)]:= \delta t^{-1}\left ( \frac{120s}{M}(n m_{12}) \right )$, that is, the class which time interval is $n \delta t [(e_1, e_2)]$ corresponds to the one characterized by the mass obtained by putting $n$ equal masses over the scale pan. 

If we define $n m_{12} : = \mu$ and remove $n_1<n$ masses of the pan, then we are left with the mass $\frac{n_2}{n} \mu$; $n_2 = n- n_1$. Hence, this prescription allows us to write    
\begin{align}\label{9.1}
	\cdot : \mathds{Q}^+ \times \mathds{V}_{\delta T} \longrightarrow \mathds{V}_{\delta T}
\end{align}   
as well. 
The next step would be to define the multiplication by real numbers. In some sense, we are limited by two peculiar problems. One of them is related to the operational prescription of measuring time intervals with the scale. It will always provide a rational value due to the experimental precision the scale possesses. We could never write $\pi s$ for a time interval for a pair of events even if our intuition says that there could exist such a pair. With \eqref{9.1} in hands, it would be possible to invoke the density of $\mathds{Q}$ in $\mathds{R}$ to construct a sequence of rationals $(q_n)$ converging to any real $q_n \rightarrow \alpha$, such that
\begin{equation}
	q_n \delta t \rightarrow \alpha \delta t
\end{equation}
but once again we are restricted by the experimental prescription. Therefore at this point it would suffice to consider only the field $\mathds{Q}$.

The other problem is related to multiplication by negative values. The negative time interval $- \delta t(e_1, e_2)$ has no meaning. We could think of withdrawing part of the mass on one pan. Now, while it is completely legitimate for the physical quantity mass, understood as removing, $-\delta t$ could only be interpreted if we admit, for example, a coordinate in the arrow of time: for example, solving for the quadratic equation when a ball will hit the ground upon being tossed up-wards gives two solutions. The positive one is the answer one is looking for. The negative solution has a meaning though. It is the answer for the time-reversal process if one is reconstructing the trajectory; in this sense, it is the time when the trajectory of the ball was intercepting the ground before the tossing time. The great volume of mathematical information and physical meaning in this last statements is not available to us up until now. 

If we ignore these two problems, then $(\mathds{V}_{\delta T}, + , \cdot)$ is an one-dimensional vector space, over the field $\mathds{R}$, as long as the two operations are well behaved in this sense. A quick experimental task confirms it. Since we won't work out differentiability or other related topics, we could have left $\mathds{Q}$ instead of $\mathds{R}$. Our construction only indicates a possible continuum line.

To conclude this Section, we outline some comments about what we have made so far. 

\textbf{1.} The first comment concerns the procedure for establishing values to time intervals. It is based on a preconceived notion of mass. We haven't presented it here and the interested reader may find it on \cite{guilherme.rodrigo.rizzuti.2018}. The main critic to it is that it is based on the local gravity, responsible for the water flow. In this case, our procedure may lead to different values for time intervals all over the Earth, even with the same procedure. Although it is a delicate obstacle, it doesn't interfere with the aim of our paper, which is mainly related to the time line completion, in the topological sense.

\textbf{2.} $(\mathds{V}_{\delta T}, + , \cdot )$ possesses the structure of a one-dimensional vector space. One of its basis is given by $\mathcal{B} = \{ 1 s\}$, which gives, in particular, an interpretation to the dimension of the physical quantity $\delta$-time. Any other time interval could also be taken as basis. Our option relies only on its common usage. We also point out that the option for looking it as an one-dimensional space was chosen by the sake of simplicity: $\mathds{R}$ is infinite dimensional over the field $\mathds{Q}$.

\begin{figure}[H]
	\centering
	\includegraphics[scale=0.4]{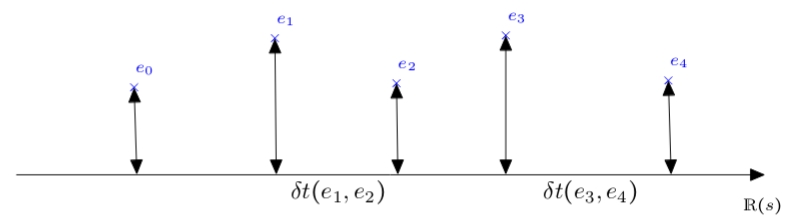}
	\caption{Exploring the isomorphism $\mathds{V}_{\delta T} \cong \mathds{R}(s)$.}
	\label{fig:aline}
\end{figure}

\textbf{3.} Two linear spaces with the same dimension are isomorphic. Since the real line $\mathds{R}$ (with the usual operations) is also one-dimensional, we could invoke such isomorphism $\mathds{V}_{\delta T} \cong \mathds{R}(s)$ to provide a coordinate over $\mathds{V}_{\delta T}$.\footnote{Throughout the paper, the notation $\mathds{R}(s)$ means the set of real numbers multiplied by the dimension of time.} Perhaps we could do even better: we ``align'' the events as in Figure \ref{fig:aline}.
Unfortunately, our operational prescription does not allow us to do that. As described before, it is not possible to measure $\delta t[(e_0, e_{\pi})] = \pi s$. Moreover, $\mathds{R}$ is equipped with a total order relation $<$. We can compare two elements of $\mathds{V}_{\delta T}$, say $\delta t_1$ and $\delta t_2$ with $<$, imported from ordering masses: $\delta t_1 < \delta t_2$, $\delta t_2 < \delta t_1$ or $\delta t_1 = \delta t_2$. However, this order does not say what events originating both intervals came first.   

\textbf{4.} Although we may say that $D_{\delta T}/ \sim_T$  is formed by classes of pairs of events, with the same time duration, that is
\begin{equation}
	(e,f), (g,h) \in [(e_1, e_2)] \Rightarrow \delta t(e,f) = \delta t (h,g),
\end{equation}
we cannot say which event came first. It would be desirable that the construction depicts the relation of future/past between a pair of events.

\textbf{5.} According to the comments above, $\mathds{V}_{\delta T}$ does not have yet the desirable (topological) ordered structure as the real line (in the sense of ordering events). However,  $\mathds{V}_{\delta T}$ plays a central role in constructing the notion of space. Roughly speaking, it is a set of points marked in a frame or system of reference in which distance of arbitrary pair of points does not change in time. The operational approach shows that the physical quantities (spatial) distance and $\delta$-time, in some sense, are necessary to construct a space $\mathscr{E}$. The distance, in particular, induces a natural compass ball-based topology in $\mathscr{E}$ \cite{rizzuti.luca.cristhiano.2019}. 

The next two comments are, in fact, severe critiques of the experimental construction of $\mathds{V}_{\delta T}$.

\textbf{6.} To use a scale to measure the lapse of time is completely dependent on the local gravity. If we explore the Einstein principle of equivalence in great extent, then, the clepsydra fails to work. For instance, you may try to measure the time lapse of a free-falling object \textit{in loco} in one of those playground falling-elevators. In this case, it is not possible to measure the time interval with a water-clock; this is precisely defined though.

\textbf{7.} The other problem concerning the elements of $\mathds{V}_{\delta T}$ is related to opening and closing the faucet. It implies the very notion of simultaneity. If it is not well defined for a pair $(e_1, e_2)$, then $\delta t [(e_1, e_2)]$ would carry an empty meaning. In our initial construction, it was implicitly assumed that it is possible to exchange information instantaneously. On the other hand, adopting an operational methodology demands an experimental prescription to accompany the corresponding concept. In our next lines, we will solve this issue related to the concept of simultaneity.    

\textbf{8.} Our operational prescription is an effort to define time using an arbitrary but suitable process. It is effectively based on the intuition that time parametrization of a process should be a symmetry, namely, the reparametrization invariance. The role of reparametrization invariance is widely discussed in \cite{gueorguiev_reparametrization_2021}, which implies, for example, a common arrow of time and non-negativity of mass. Broader consequences, such as long range interactions (electrodynamics and gravity) may also be justified by reparametrization-invariant systems, see \cite{gueorguiev_geometric_2021}. This idea of considering time as a symmetry has also been exploited in quantum mechanics, see \cite{aharonov.b.l}.  

Some of the difficulties raised in the comments above can be overcome with the help of some geometry of free particles and a couple of idealized assumptions. Moreover, this will also answer the question the title of the paper proposes to attack.

\section{Free particles as clocks}

Following an operational paradigm, we will reformulate the pragmatic construction of the initial part of this work into a more precise though abstract proposal equivalent to the initial one. Besides that, we will also shed light on the problems exposed in the comments in the previous Section.  

As discussed in \cite{rizzuti.luca.cristhiano.2019}, the concept of space is related to the existence in nature of rigid bodies, that is, objects whose distance between arbitrary pair of points does not change with time. They may be glued together to form another rigid body. Such union is an equivalence relation in the set of all rigid bodies and we name it frame or system of reference (SR). Finally, the space $\mathscr{E}_{SR}$ associated to a SR is the set of points that could have, and are marked in this way.

One of our main objectives with this work is to study the topological structure of the arrow of time. We can explore it with the very notion of free particles and their corresponding trajectories, as we shall see later. It has a close connection to our clepsydra, to be explained in a while. Moreover, this type of geometrization could fill the blanks raised on the previous Section. By particle, we understand an object whose characteristic size is much smaller than the distances involved in the corresponding description: it can be localized at a point in space. Our daily experience shows that some particles have simpler movements than others. Moreover, we may interfere with the movement, by exerting forces on the particles. 

We can, however, observe and prepare particles so far away from any external interference.\footnote{Of course it is impossible to remove gravitational effects. Nevertheless, a falling particle, is, at least locally, free, due to the principle of equivalence.} They will be called free particles. Furthermore, we take another idealized assumption that there exist systems of reference where the trajectory of any free particle is a point or a straight line. They will be called inertial systems of reference. There are two observations needed. The first one is related to the uniformity between distances traveled by two distinct free particles. If one of the particles travels the distance $d_1$, while the other $d_2$, the ratio between distances will be always $d_1/d_2$, no matter the time interval. In this sense, we could forget about time as a parameter of evolution and instead, use a standard particle and its displacement to do the task. The second observation brings the water-clock to the game. We prepare a free particle in an inertial system of reference to walk through the straight line $s$. In $s$, we mark equal distanced points $E_1$, $E_2$,..., $E_n$, that is, $d(E_1, E_2) = d(E_2, E_3) = \cdot \cdot \cdot = d(E_{n-1}, E_n)$ for some natural number $n \in \mathds{N}$. Consider the events  $e_1$, $e_2$,..., $e_n$ consisting in the arriving of the particle at each corresponding point. We have the following experimental facts, 
\begin{align}\label{16.1}
	\frac{d(E_1, E_k)}{d(E_1, E_2)} = \frac{\delta t(e_1, e_k)}{\delta t (e_1, e_2)}, \, \forall k \in \{1,...,n \}. \\ \label{16.2} 
	(e_k, e_{k+1}) \in [(e_1, e_2)], \, \forall k \in \{1,...,n-1 \}.
\end{align}
Let us enumerate some of their consequences:
\begin{enumerate}
	\item It is exactly this kind of uniformity in \eqref{16.1} that assures the equivalence between the geometrical (time) evolution given by the ratio $d_1/d_2$ and ours, with the clepsydra. 
	
	\item Putting aside the precision of the clepsydra and our capacity to mark arbitrarily close distanced points in $s$, we generate a sequence $(E_n) \subset \mathscr{E}_{SR}$ and $(e_n) \subset \mathscr{ET}$. This means, in particular, that for any free particle in any inertial system of reference (such that its trajectory is not a point),
	\begin{enumerate}[label=\alph*)]
		\item For any event of arrival, the particle will be in only one point;
		
		\item The particle will be at any point in its trajectory in no more than one event.
		
	\end{enumerate}
	In other words, there is a bijection between $(E_n)$ and $(e_n)$.
	
	\item With this bijection in hands, there is a natural way to define a (temporal) partial order of events. We construct the displacement vector $\overrightarrow{E_1 E_2}$ and then, one defines a total order of points $(E_n) \subset s$. Given $P$ and $F$ in $(E_n)$, we say that
	\begin{equation}\label{lambda}
		P < F \Leftrightarrow \exists \lambda \in \mathds{R}^+_*; \, \overrightarrow{PF} = \lambda \overrightarrow{E_1 E_2}.
	\end{equation} 
	This paves the way to define a temporal order of events. We say that the event $p$ is in the past of $f$ ($p \prec f$), or conversely, that $f$ is the future of $p$ ($f \succ p$) whenever $P < F$. $p$ and $f$ are elements of $(e_n)$ in one-to-one correspondence with $P$ and $F$. The reason this order relation is partial and not total is that, given some $SR$ and events $e, f \in \mathscr{E}_{SR}$, we need to find a free particle that experiences both $e$ and $f$. The problem is that there are situations where this cannot be achieved, namely when $e$ is \textbf{simultaneous} to $f$. This word will be properly defined in the next section.
	
	The careful reader will notice a small trick here. The expression \eqref{lambda} requires that space is simply connected. Since our analysis is based upon the notion of free particles, we could use, for example, the conservation of mass to make \eqref{lambda} consistent in this sense. This way, we avoid 
	that the particle could disappear, leaving \eqref{lambda} with no significance whatsoever. We stress that our work has a classical background. It prevents awkward situations, such as creation and annihilation of particles, likely to happen in the quantum realm. It also circumvents the very notion of anti-particles, where time order could be inverted.    
\end{enumerate}

\section{The problem of simultaneity}

The problem of simultaneity has not been addressed yet. This is critical at this stage as one would say that not all pairs of events have a defined time interval. If the pair may be ``connected'' by a free particle, then the corresponding time interval is well-defined. We point out that there is no superior bound for the factor $\frac{d(E_1, E_2)}{\delta t (e_1, e_2)}$ given in \eqref{16.1}, since relativity was not taken into account. Actually, when one postulates the speed of light as a superior invariant scale, the notion of simultaneity is reformulated and, in some sense, its relativistic meaning is simpler than the corresponding Newtonian counterpart. Let us discuss the latter in detail, while we direct the reader to the usual literature \cite{urbantke} for the former.

Returning now to an arbitrary pair of events $e_1$ and $e_2$, it may happen that $e_1$ and  $e_2$ take place ``at the same time''. These words are too vague to be used. We have an idea of what they mean though. If $e_1$ and $e_2$ happen at the same time, clearly there is a natural spatial distance between the corresponding points $E_1$ and $E_2$ where it takes place. After all, the experimental prescription of measuring $d(E_1, E_2)$ consists of setting the compass or ruler simultaneously  on $E_1$ and $E_2$. That's the hint we needed: we use geometric configurations to define the notion of simultaneity. We will see that with this construction, the entire $\mathscr{ET}$ is unified by one arrow of time. There is a simple heuristic example that explains the spark just given above \cite{site.df.bernhard}.

\begin{figure}
	\centering
	\includegraphics[scale=0.23]{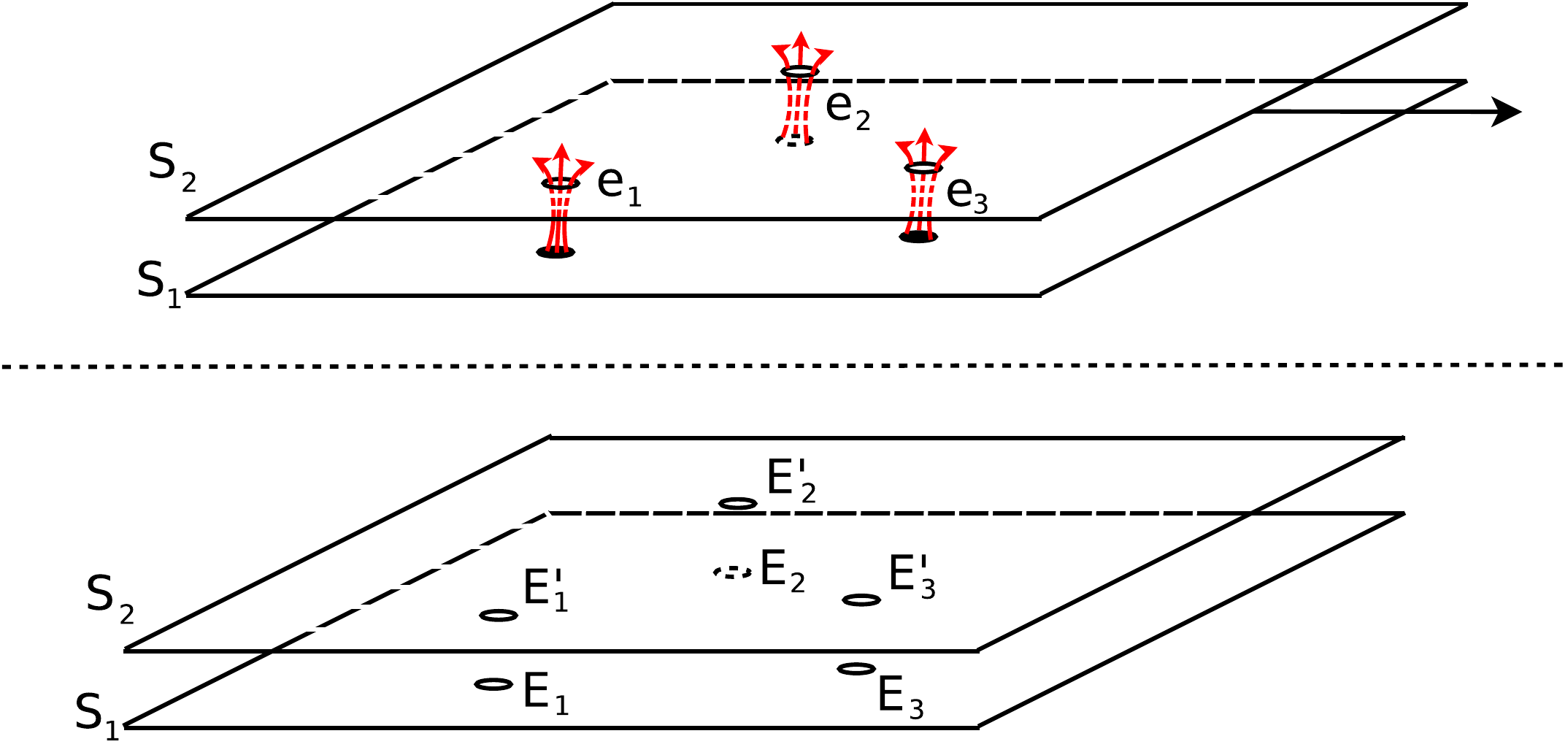}
	\caption{Example of simultaneous events.}
	\label{explosao}
\end{figure}

\begin{figure}
	\centering
	\includegraphics[scale=0.23]{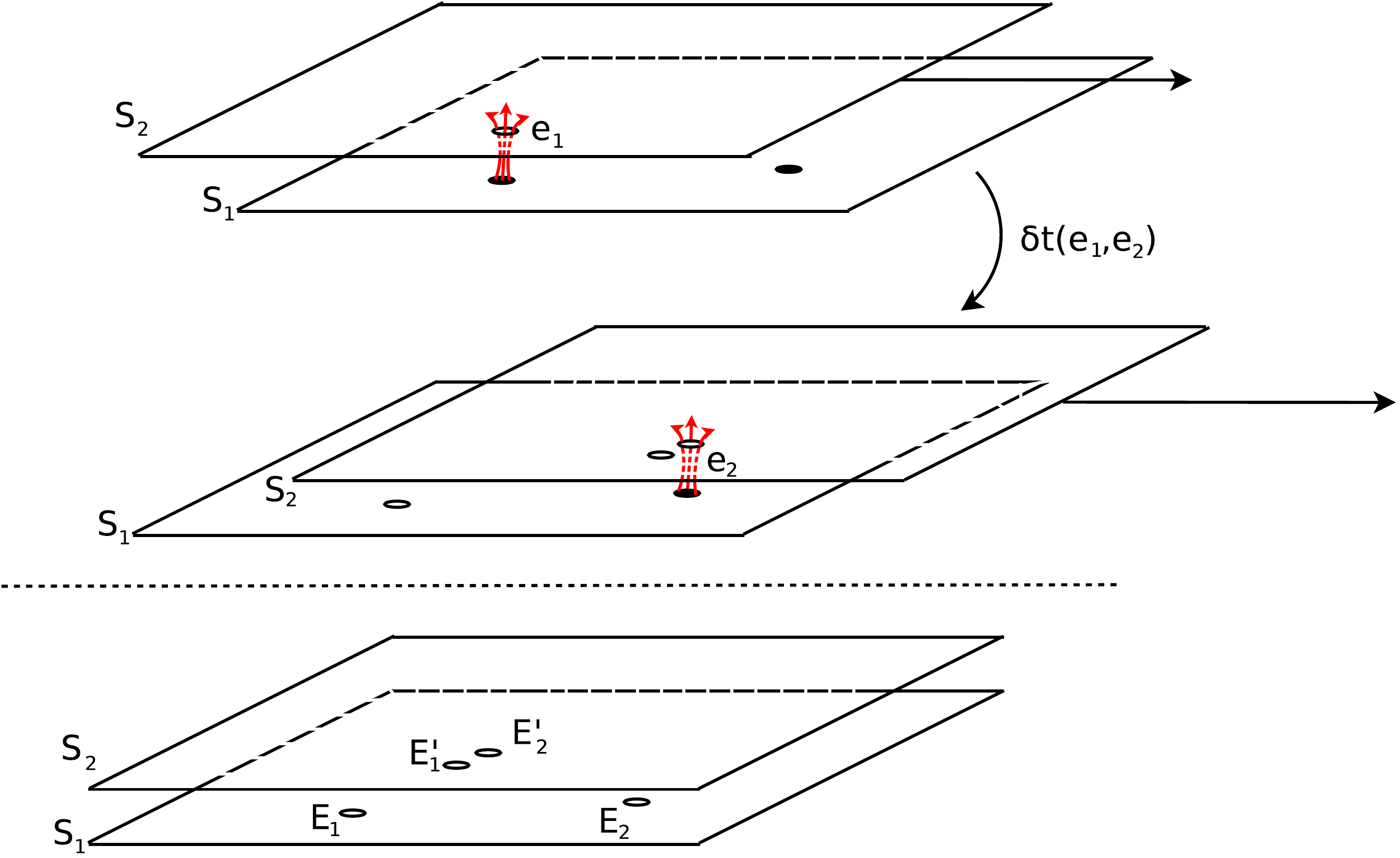}
	\caption{Example of non-simultaneous events.}
	\label{explosao_2}
\end{figure}

The idea of the intuitive example resides in the comparison of geometric relations between points defined in different systems of reference by some set of events. So, let us consider a set of $n$ little gunpowder mounds in a sheet of paper. Without deformations (bending, for instance), the sheet may be considered a system of reference. The mounds are connected to an electric startup device that lights the gunpowder and makes them explode, generating some events, say, $e_1$,..., $e_n$. We let another sheet of paper fly parallel over the first one. The startup gear is fired and we compare the distance between points $E_1$,..., $E_n$ marked in the first and points $E'_1$,..., $E'_n$ marked in the second sheet of paper due to the little explosions. We say that the events are simultaneous whenever the figures defined by the corresponding points in both systems are the same. That is, distances are preserved. Figure \ref{explosao} depicts this example. In the first part, $S_2$ flies over $S_1$ and the three little mounds explode, generating the events $e_1$, $e_2$ and $e_3$. The second part of the Fig. \ref{explosao} compares the distances between the corresponding points, in the different systems of reference. In this case, $d(E_i, E_j) = d(E'_i, E'_j)$, for $i \neq j \in \{1, 2, 3\}$. 
On the other hand, Figure \ref{explosao_2} shows the case where the events $e_1$ and $e_2$ are not simultaneous, as the distances between the points so created in each system is clearly different, $d(E_1, E_2) \neq d(E'_1, E'_2)$. Observe the time lapse between $e_1$ and $e_2$. 

Let us formalize this na\"ive example. For each event in the space-time, we associate it to a point in a space of a system of reference. So, there exists a map, that we name localization \cite{site.df.bernhard},
\begin{align}
	L_{SR}: \mathscr{ET} &\longrightarrow \mathscr{E}_{SR} \cr
	e &\longmapsto L_{SR}(e).
\end{align}

On the other hand, \underline{for each point} $P \in \mathscr{E}_{SR}$, we may choose an event $e$ that happens in the point $P$, defining another map 
\begin{align}
	\Omega_{SR}: \mathscr{E}_{SR} &\longrightarrow \mathscr{ET} \cr
	P &\longmapsto \Omega_{SR}(P).
\end{align}
We point out that

(i) $L_{SR}$ is not invertible in general as different events may happen at the same point.

(ii) The comment above explains why we have used the words ``\emph{an} event $e$, for \emph{each} point $P$''.  In this case, $\Omega_{SR}$ is a well-defined map. The emphasized words were not expendable.

In light of the observations above, we have the following
\begin{definition}
	Two events $e_1$ and $e_2$ are \textbf{simultaneous} when the distance between the pair $(L_{SR}(e_1)$, $L_{SR}(e_2))$ is independent of the reference system.
\end{definition}
To elucidate this definition further, we can use our example with the gunpowder mounds. The static and moving pieces of paper represent each some system of reference and the markings left on them after the explosions are the images of their respective localization maps. Note that we could have the piece of paper moving at any speed, and the markings would still be the same as in the static piece of paper in the first example. However, in the second, depending on the speed we would measure different distances between the markings. In particular, using an appropriate speed we could move a piece of paper in such a way to make the markings at the same point.

Simultaneous events can also be characterized by

\begin{proposition}\label{simnopart}
	Two distinct events are simultaneous if and only if no free particle experiences both.\footnote{Equivalently: two events connected by a free-particle are not simultaneous.}
\end{proposition}
\textit{Proof.} $(\Rightarrow)$ Let $e_1$ and $e_2$ be two distinct simultaneous events, that is, $d(E_1, E_2)>0$ is independent of the frame. Suppose that there is a particle which experiences both events. In the frame of the particle, the distance between the corresponding pair of points would be $0$, contradicting our hypothesis.

$(\Leftarrow)$ In this case, we directly show the contrapositive. Let $e_1$ and $e_2$ be two non simultaneous events. It means that the distance $d(L_{SR}(e_1)$, $L_{SR}(e_2))$ is frame dependent. We thus use the clepsydra to find $\delta t(e_1, e_2)$ and prepare a free particle with velocity 
$$\vec{V} = \frac{\overrightarrow{L_{SR}(e_1) L_{SR} (e_2)}}{\delta t(e_1, e_2)}.$$
By construction, this particle experiences both $e_1$ and $e_2$.

\qed

\begin{cor}\label{corpartsim}
	Given $e$ and $f$ two non simultaneous events, then either $e \prec f$ or $f \prec e$.
\end{cor}
\textit{Proof.} If $e$ and $f$ are non simultaneous, then there is a particle experiencing both of them. Therefore we can apply our definition for the past-future order described previously. 

\qed

An important consequence of the definition of simultaneity is the
\begin{proposition}
	Simultaneity is an equivalence relation.
\end{proposition}
\textit{Proof.} Let us denote the relation of simultaneity by $S$.
\begin{enumerate}[label=(\roman*)]
	\item For any system of reference $SR$ and any event $e_1$, $d(L_{SR}(e_1), L_{SR}(e_1)) = 0$, that is, $e_1 S e_1$;
	
	\item If $e_1 S e_2$, then $d(L_{SR}(e_1), L_{SR}(e_2))$ for any system of reference $SR$. But as $d$ is symmetric, $d(L_{SR}(e_1), L_{SR}(e_2)) = d(L_{SR}(e_1), L_{SR}(e_2))$, for any $SR$. Therefore, $e_2 S e_1$.
	
	\item To conclude that $S$ is transitive, we use an old fact from geometry: given three points, say $E_1$, $E_2$ and $E_3$ and the distances $d(E_1, E_2)$ and $d(E_2, E_3)$, then $d(E_1, E_3)$ is uniquely determined. So, consider an arbitrary $SR$ and events $e_1$, $e_2$ and $e_3$ such that $e_1 S e_2$ and $e_2 S e_3$. By definition, the distances $d(E_1, E_2)$ and $d(E_2, E_3)$ are well defined and are the same, for any $SR$. Hence, in light of the what was initially stated, the distance $d(E_1, E_3)$ is fixed and independent of $SR$, that is, $e_1 S e_3$.
\end{enumerate}
\qed

\begin{figure}
	\centering
	\includegraphics[scale=0.3]{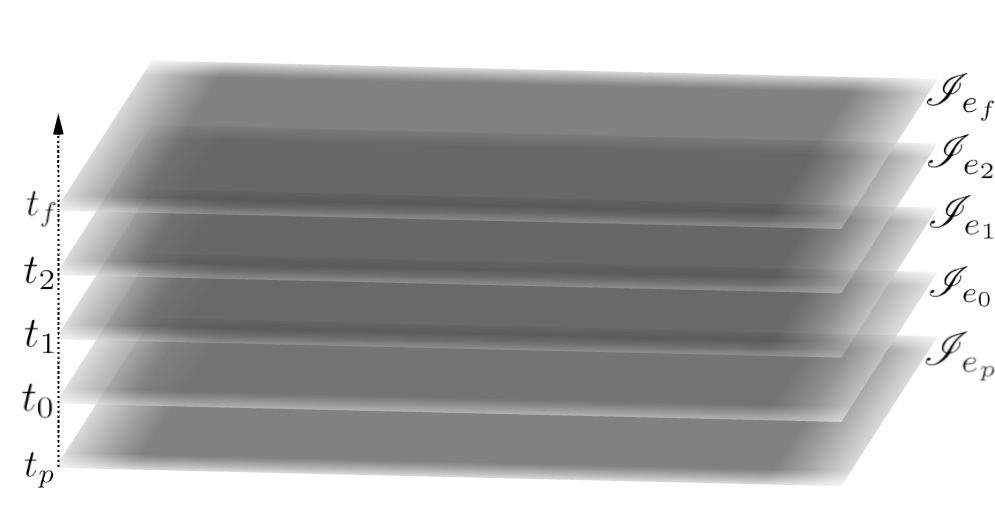}
	\caption{Universal arrow of time.}
	\label{arrow}
\end{figure}

We will denote the equivalence class under simultaneity of an event $e_1$ by $\mathscr{I}_{e_1}$ and will call an \textbf{instant} (when $e_1$ happens).  The word instant has already been used before, see Definition \ref{definition.01}. Each class of simultaneous events will be accommodated in one, and only one, point in the arrow of time, as we will see in a while (see Fig. \ref{arrow}). Hence, it justifies the connection of the former popular usage to latter formal one just defined. Moving on, one important property of these classes is the

\begin{proposition}
	Given a system of reference $SR$ and an instant $e_0$, the restriction $L_{SR}\vert_{\mathscr{I}_{e_0}}$ is a bijection.
\end{proposition}
\textit{Proof.} Let $e_1 \neq e_2$, such that $E_i :=L_{SR}(e_i), i=1,2$. Since no particle can experience both, it would never happen $d(L_{SR}(e_1), L_{SR}(e_2))=0$ in any frame. Thus, $d(E_1, E_2)>0$, which implies $E_1 \neq E_2$, as $d(\cdot, \cdot)$ is a metric. Equivalently, $L_{SR}(e_1)\neq L_{SR}(e_2)$. It shows the injectivity. For surjectivity, we take $E \in L_{SR}\vert_{\mathscr{I}_{e_0}}$. The distance $d(E, E_0)$ is well defined. Owing to our intuitive example in Figures \ref{explosao} and \ref{explosao_2}. $E$ and $E_0$ could be thought as two simultaneous events of the explosions $e$ and $e_0$. This way, given any point $E$ in $L_{SR}\vert_{\mathscr{I}_{e_0}}$, we may always find an event $e$, such that $L_{SR}(e) = E$.

\qed

This proposition gives us a way to define a metric on the instant $\mathscr{I}_{e_0}$.
\begin{definition}
	We define the \textbf{instantaneous distance} in the instant of $e_0$ as the map $d^{e_0}: \mathscr{I}_{e_0} \times \mathscr{I}_{e_0} \rightarrow \mathds{R}(m)$, given by $(e_1, e_2) \mapsto d(L_{SR}(e_1), L_{SR}(e_2))$.
\end{definition}
Note that we are able to define the map above because the distance between the localization of events in the same instant is independent of the system of reference.
\begin{proposition}
	Given an event $e_0$, $d^{e_0}$ is a metric on $\mathscr{I}_{e_0}$.
\end{proposition}
\textit{Proof.}
All the metric properties come from the fact that $d$ is a metric.
\qed

The proposition above guarantees that the restriction of the localization map has an inverse, which is the map $\Omega_{SR}^{e_0}: \mathscr{E}_{SR} \rightarrow \mathscr{I}_{e_0}$. Note that this is the restriction of $\Omega_{SR}$ defined before, to the instant $\mathscr{I}_{e_0}$. In particular, $\Omega_{SR}^{e_0}$ is a bijection for any event $e_0$, as it has an inverse. With this remark we can now prove the
\begin{theorem}\label{theo1}
	Given two systems of reference $SR$ and $SR'$, the map $L_{SR} \circ \Omega_{SR'}^{e_0}$ is a bijective isometry.
\end{theorem}
\textit{Proof.}
Given an event $e_0$, the metric on $\mathscr{I}_{e_0}$ is defined in such a way to make the map $L_{SR}\vert_{\mathscr{I}_{e_0}}$ distance preserving, for any system of reference $SR$. Now, let $e_1, e_2 \in \mathscr{I}_{e_0}$ be events such that $L_{SR}(e_1) = P$ and $L_{SR}(e_2) = Q$. In this case, we have
\begin{align}
	d^{e_0}(\Omega^{e_0}_{SR}(P), \Omega^{e_0}_{SR}(Q)) = d^{e_0}(e_1, e_2) = d(L_{SR}(e_1), L_{SR}(e_2)) = d(P, Q),
\end{align}
i.e., $\Omega^{e_0}_{SR}$ is distance preserving for any system of reference $SR$. Then, given system of reference $SR$ and $SR'$, the composition
\begin{align}
	L_{SR'} \circ \Omega_{SR}^{e_0}: \mathscr{E}_{SR} \rightarrow \mathscr{E}_{SR'}
\end{align}
is distance preserving as it is the composition of two distance preserving maps. Moreover, it is the composition of bijective maps, meaning itself is bijective.
\qed

\begin{theorem}\label{theo2}
	For any system of reference $SR$ and any events $e_1, e_2 \in \mathscr{ET}$, $\Omega_{SR}^{e_2} \circ L_{SR}\vert_{\mathscr{I}_{e_1}}: \mathscr{I}_{e_1} \rightarrow \mathscr{I}_{e_2}$ is an isometry.
\end{theorem}
\textit{Proof.} As already observed before, for any $SR$, $L_{SR}\vert_{\mathscr{I}_{e_1}}$ and $\Omega_{SR}^{e_2}$ are bijections. Moreover, as pointed out in the proof of $\ref{theo1}$, they are isometries for any events and system of reference. Thus, the composition $\Omega_{SR}^{e_2} \circ L_{SR}\vert_{\mathscr{I}_{e_1}}$ is also a bijective isometry.

\qed

We now generalize the definition of $\delta t$ to instants. It doesn't make sense to speak of time-lapse for a pair of events that don't occur in the same point in space, but for a pair of instants. Since any event in $\mathscr{I}_{e_1}(\mathscr{I}_{e_2})$ is simultaneous with $e_1 (e_2)$, we write, instead of $\delta t(e_1, e_2)$ for a specific pair or $\delta t[(e_1, e_2)]$ for a class,
\begin{equation}
	\delta t (\mathscr{I}_{e_1},\mathscr{I}_{e_2}).
\end{equation}
The process of measuring $\delta t$ is made by choosing events $e_1'$ in $\mathscr{I}_{e_1}$ and $e_2'$ $\mathscr{I}_{e_2}$ and measuring $\delta t(e_1', e_2')$ as we show in
\begin{proposition}\label{timedistinstants}
	$\delta t (\mathscr{I}_{e_1},\mathscr{I}_{e_2})$ is independent of the events $e_1 \in \mathscr{I}_{e_1}$ and $e_2 \in \mathscr{I}_{e_2}$. 
\end{proposition}
\noindent
\textit{Proof.} Let $e_1 \neq e'_1$ be arbitrary events in $\mathscr{I}_{e_1}$ and $e_2 \neq e'_2$ in $\mathscr{I}_{e_2}$. To measure the corresponding time-lapses, the faucet is open when $e_1$ or $e'_1$ happens and is closed when $e_2$ or $e'_2$ happens. So, we have
\begin{equation}\label{24.1}
	\delta t(e_1, e_2) = \delta t(e'_1, e'_2) = \delta t (\mathscr{I}_{e_1}, \mathscr{I}_{e_2}).
\end{equation}
On the other hand, if one uses free particles to measure the corresponding time-lapse, they must be prepared with different velocities, depending on the spatial distance of the respective points in space where the pair of events takes place. If $d(E_1, E_2)$ is greater than $d(E'_1, E'_2)$, the velocity of the free particle is adjusted accordingly so that the time-lapse \eqref{24.1} will remain the same.

\qed

Theorem \ref{theo2} implies an interesting fact about instants. Given distinct instants $\mathscr{I}_{e_1}$ and $\mathscr{I}_{e_2}$, as $e_1$ and $e_2$ are not simultaneous by Corollary \ref{corpartsim} either $e_1 \prec e_2$ or $e_2 \prec e_1$. Without loss of generality, let us suppose the first case. 
Then, we say that $\mathscr{I}_{e_1} \preccurlyeq \mathscr{I}_{e_2}$, defining a total order relation in the set of instants.\footnote{This order will be indeed total, as the partial order of events was only not defined for events in the same instant. When we collapse all the simultaneous events in the same class, this does not occur, and so the order becomes total.} To show that this is well defined, we must show that it is independent of the class representative. Let us take $e \in \mathscr{I}_{e_1}$ and $f \in \mathscr{I}_{e_2}$. Suppose we had $f \prec e$. Then, by Proposition \ref{simnopart} there is a free particle experiencing both $e_1$ and $f$ and another which experiences both $f$ and $e$. However, taking a particle moving with the sum of the velocities of these two particles, there would be a free particle between $e$ and $e_1$ contradicting Proposition \ref{simnopart}.

Now that we have shown that $\delta t$ is well-defined, we can prove
\begin{proposition}\label{quasimetric}
	For any pair of instants $\mathscr{I}_{e_1}$ and $\mathscr{I}_{e_2}$ $\delta t$ has the following properties:
	\begin{enumerate}[label=(\roman*)]
		\item $\delta t(\mathscr{I}_{e_1}, \mathscr{I}_{e_2}) \geq 0$;
		
		\item $\delta t(\mathscr{I}_{e_1}, \mathscr{I}_{e_2}) = 0 \Leftrightarrow \mathscr{I}_{e_1} = \mathscr{I}_{e_2}$;
		
		\item for any $\mathscr{I}_{e_3}$, $\delta t(\mathscr{I}_{e_1}, \mathscr{I}_{e_2}) \leq \delta t(\mathscr{I}_{e_1}, \mathscr{I}_{e_3}) + \delta t(\mathscr{I}_{e_3}, \mathscr{I}_{e_2})$.
	\end{enumerate}
\end{proposition}
\noindent
\textit{Proof.} Let $\mathscr{I}_{e_1}$ and $\mathscr{I}_{e_2}$ be a pair of instants.
\begin{enumerate}[label=(\roman*)]
	\item To measure $\delta(\mathscr{I}_{e_1}, \mathscr{I}_{e_2})$ we open and close our faucet when $e_1$ and $e_2$ happen, respectively. As the weight of water is always a positive number (defined in \cite{guilherme.rodrigo.rizzuti.2018}), we have that $\delta(\mathscr{I}_{e_1}, \mathscr{I}_{e_2}) \geq 0$.
	
	\item If $\delta t(\mathscr{I}_{e_1}, \mathscr{I}_{e_2}) = 0$, this means no water flowed between the moment we opened and closed our faucet, which happens if, and only if $e_1$ and $e_2$ happen in the same instant, that is, $\mathscr{I}_{e_1} = \mathscr{I}_{e_2}$.
	
	\item Given a third instant $\mathscr{I}_{e_3}$ and supposing, without loss of generality, that $\mathscr{I}_{e_1} \preccurlyeq \mathscr{I}_{e_2}$, there are three self-excluding possibilities: $\mathscr{I}_{e_1} \preccurlyeq \mathscr{I}_{e_3} \preccurlyeq \mathscr{I}_{e_2}$, $\mathscr{I}_{e_3} \preccurlyeq \mathscr{I}_{e_1} \preccurlyeq \mathscr{I}_{e_2}$  or $\mathscr{I}_{e_1} \preccurlyeq \mathscr{I}_{e_2} \preccurlyeq \mathscr{I}_{e_3}$. By a simple experiment, which mainly consists of comparing masses delivered by the faucet in the clepsydra, we can determine that
	\begin{align}
		\delta t(\mathscr{I}_{e_1}, \mathscr{I}_{e_2}) = \delta t(\mathscr{I}_{e_1}, \mathscr{I}_{e_3}) + \delta t(\mathscr{I}_{e_1}, \mathscr{I}_{e_3})
	\end{align}
	in the first case and
	\begin{align}
		\delta t(\mathscr{I}_{e_1}, \mathscr{I}_{e_2}) < \delta t(\mathscr{I}_{e_1}, \mathscr{I}_{e_3}) + \delta t(\mathscr{I}_{e_1}, \mathscr{I}_{e_3})
	\end{align}
	in the second and third cases. Combining these two results we obtain for any instant $\mathscr{I}_{e_3}$
	\begin{align}
		\delta t(\mathscr{I}_{e_1}, \mathscr{I}_{e_2}) \leq \delta t(\mathscr{I}_{e_1}, \mathscr{I}_{e_3}) + \delta t(\mathscr{I}_{e_1}, \mathscr{I}_{e_3})
	\end{align}
	usually known as the triangle inequality.
	
\end{enumerate}
\qed

Proposition \ref{quasimetric} gives us three of the four necessary properties for $\delta t$ to be a metric. The fourth one needed is reflexivity, that is, $\delta t(\mathscr{I}_{e_1}, \mathscr{I}_{e_2}) = \delta t(\mathscr{I}_{e_2}, \mathscr{I}_{e_1})$. Now, our definition of $\delta t$ does not allow us to measure $\delta t(\mathscr{I}_{e_2}, \mathscr{I}_{e_1})$ at all, as $e_1$ happens before $e_2$. This means that we would first close the faucet and then open it, thus giving us no water. Could we fix this problem in some way? Maybe this is more a matter of notation than anything else. Let us define $\delta t(\mathscr{I}_{e_1}, \mathscr{I}_{e_2})$ as the weight of water that flows by opening the faucet when \textbf{either} $e_1$ or $e_2$ happens and close it when the other happens. By defining $\delta t$ in this way, we are simply defining that $\delta t$ is symmetric. Although this may seem  like we are cheating, this definition represents our intuitive idea that if time were to flow backwards, the amount of water measured would be the same.

For the readers who still feel uncomfortable with our ``fix'' above, it is still possible to define a symmetric $\delta t$ in terms of free particles. The intuitive idea would be to measure the ``length'' of the trajectory of a particle that goes through events $e_1$ and $e_2$. This formalizes somewhat the idea of time flowing backwards, but the measurement of this ``length'' would necessitate a greater amount of mathematical tools and will not be explored here. Nevertheless, it seems to be a fine assumption to say that $\delta t$ is symmetric.

With this result in hands, we have
\begin{cor}
	Given events $e_1 \prec e_2 \prec e_3$
	\begin{equation}\label{24.2}
		\delta t (\mathscr{I}_{e_1},\mathscr{I}_{e_2}) + \delta t (\mathscr{I}_{e_2},\mathscr{I}_{e_3}) := \delta t (\mathscr{I}_{e_1},\mathscr{I}_{e_3}).
	\end{equation}
\end{cor}
This is a result from the proof of Proposition \ref{quasimetric} item (iii).

The watchful reader observes here a remarkable fact, completely analogous to the geometric sum of distances. For arbitrary points $A$, $B$ and $C$ in space, one has
\begin{equation}\label{r.1}
	d(A,C) \leq d(A,B) + d(B,C).
\end{equation}
The equality holds in a particular geometric setting of the points. For
\begin{equation}\label{r.2}
	d(A,C) = d(A,B) + d(B,C) 
\end{equation}
we say that the points are aligned and this was used to define a straight line \cite{rizzuti.luca.cristhiano.2019}.

Due to the past-future dichotomy expressed before, space-time is sliced by the subsets $\mathscr{I}$ of simultaneous events. There is a natural label for each $\mathscr{I}$, which is tagged according to the following prescription. We choose an arbitrary event $e_0$ and obtain $\mathscr{I}_{e_0}$. Thus, one defines the label $t_0:= 0$ to $\mathscr{I}_{e_0}$. Now, given $e_1$, we set $t_1:= \delta t (\mathscr{I}_{e_0}, \mathscr{I}_{e_1})$, if $e_1 \succ e_ 0$. Else, if $e_1 \prec e_0$, then  $t_1:= -\delta t (\mathscr{I}_{e_1}, \mathscr{I}_{e_0})$. The labels given in this way, together with the comment around equations \eqref{r.1} and  \eqref{r.2}, allow us to picture a unifying arrow of time, as shown in Figure \ref{arrow}.

Actually, Figure \ref{arrow} resumes our entire work. Space-time admits a label for each set of simultaneous events called instant. The instants are aligned in temporal order.  The temporal distance of a pair of events is well defined and can be measured with a clepsydra. It gives, as a result, a rational number with the corresponding unit, forming the set $\mathds{Q}(s)$. The transition to the continuous may be concluded by the following remark. If the temporal distance is measured by a free particle, admitting that the particle must be in every point of its trajectory (that is a straight line) for each instant, then exploring the bijection of points and events for the particle, we may conclude that the dashed line in Figure \ref{arrow} may be completed to form the set $\mathds{R}(s)$. Mathematically this can be done since the set of instants is a metric space \cite{sen.2010}. We can find an isometric immersion, $i: \mathds{Q}(s) \rightarrow \mathds{R}$ by taking each label in $\mathds{Q}(s)$ and associating it to its respective rational number, that is $\mathds{Q}(s) \ni qs \mapsto q \in \mathds{Q}$. This map is clearly injective and is dense in $\mathds{R}$. Moreover, we have
\begin{align}
	\delta t(i(\mathscr{I}_{e_1}), i(\mathscr{I}_{e_2})) = \vert i(q_1 s) - i(q_2 s)\vert = \vert q_1 - q_2\vert = d(q_1, q_2),
\end{align}
that is, $i$ is an isometric immersion. We also know that this completion is unique up to isometry, meaning that we are allowed to give the completion the name $\mathds{R}(s)$ as stated before.

\section{The time interval topology uncovered}\label{sec5}

One consequence of our work is that the operational construction of the arrow of time also unveils the topological anatomy of space-time and how it is connected to the corresponding causal structure. Let us expand it a little further. 

We choose initially an event $e_1$, obtain $\mathscr{I}_{e_1}$ defined by the label $t_1$. All the events which are in the future of $e_1$ belong to the family of instants $(\mathscr{I}_{e_t})_{t>t_1}$, with $t>t_1$. They are in the set
\begin{equation}
	\mathscr{F}(e_1):= \{ e_t \in \mathscr{ET}\, \vert \, e_t \in \mathscr{I}_{e_t}; \, t>t_1 \}.
\end{equation}
Accordingly, the past of the event $e_1$ shall be defined as
\begin{equation}
	\mathscr{P}(e_1):= \{ e_t \in \mathscr{ET} \, \vert \, e_t \in \mathscr{I}_{e_t}; \, t<t_1 \}.
\end{equation}    

\begin{figure} [H] 
	\centering
	\includegraphics[scale=0.3]{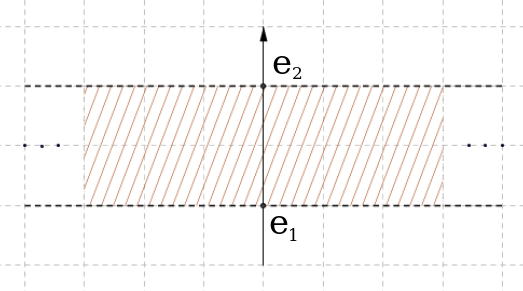}
	\caption{Geometrical scheme for $I(e_1, e_2)$.}
	\label{openset_1}
\end{figure}

We are now ready to define what will be our open sets. Let $e_1 \prec e_2$, that is $e_1 \in \mathscr{P}(e_2)$ and $e_2 \in \mathscr{F}(e_1)$. This way, 
\begin{definition}
	The set
	\begin{equation}\label{23}
		I(e_1, e_2):= \mathscr{F}(e_1) \cap \mathscr{P}(e_2)  
	\end{equation}
	is called an \textbf{open interval}. 
\end{definition}
The Fig. \ref{openset_1} exhibits a geometrical representation of $I(e_1, e_2)$. 
The interval so defined is non-empty. In fact, for any $\mathscr{I}_{e_t}$, with $t_1< t < t_2$, $\mathscr{I}_{e_t} \subset I(e_1, I_2)$. The continuum obtained previously allows us to choose such $t$. Actually, even if we haven't made the transition from $\mathds{Q}$ to $\mathds{R}$, we could choose such $t$, due to the density of the former in the latter. 
\begin{proposition}
	The set
	\begin{equation*}
		\mathcal{B}= \{ I \subset \mathscr{ET}\, \vert \,\,\, I \mbox{is an interval} \}    
	\end{equation*}
	is a basis for a topology in $\mathscr{ET}$.
\end{proposition}
\noindent
\textit{Proof.} Let $e_t$ be an arbitrary space-time event. Take $t_1 < t < t_2$ and choose events $e_1 \in \mathscr{I}_{t_1}$ and $e_2 \in \mathscr{I}_{t_2}$. By construction, $e_t \in I(e_1, e_2)$.

Moreover, let $I_a$ and $I_b$ be intervals with $I_a \cap I_b \neq \emptyset$. By assumption, there exist $t_1< t_2 < t_3 < t_4$ with corresponding instants $\mathscr{I}_{t_i}$, $i=1,2,3,4$, such that $I_a=I(e_1, e_3)$ and $I_b=I(e_2, e_4)$. $e_i$, $i=1,2,3,4$ are arbitrary events in the corresponding instants. Let $e_{t^*} \in I_a \cap I_b$. It means that $t_2 < t^* < t_3$. Consider now
\begin{equation*}
	t_m = \min \{ t^*- t_2, t_3 - t^* \}. 
\end{equation*}
We construct the instants $\mathscr{I}_{t_{\pm}}$, with $t_{\pm}:= t^* \pm t_m/2$ and take $e_{\pm} \in \mathscr{I}_{t_{\pm}}$. Define $I_c = I(e_-,e_+)$. The chain 
$$ t_2< t_-< t^* < t_+ < t_3 $$
allows us to conclude that $e_t \in I_c \subset I_a \cap I_b$. 

\qed 

So, one defines the time-time interval topology
\begin{equation}
	\tau_{\mathcal{B}} = \{ \mathcal{A}\subset \mathscr{ET}\, \vert \, \forall e \in \mathcal{A}, \, \exists I \in \mathcal{B}; \, e \in I \subset \mathcal{A} \}
\end{equation}
and the intervals are the open sets. 

The curious fact about this topology is that
\begin{proposition}
	$\tau_{\mathcal{B}}$ is not Hausdorff. 
\end{proposition}
\noindent
\textit{Proof.} Let $e_1$ and $e_2$ be arbitrary events in an instant $\mathscr{I}_t$. Every interval that contains $e_1$ will also contain $e_2$. Thus, they cannot be housed off by non-intersecting intervals.

\qed

The quotient topology (with $\mathcal{B}$) in the space of instants is the same of the real line (or $\mathds{Q}$). Our construction is consistent with the usual mathematical methods to find a topology in the quotient space. In fact, in our case, the quotient space is constructed upon the simultaneity equivalence relation. To obtain the topology on the arrow of time, we use the following standard prescription: a set in $\mathds{R}(s)$ is open whenever its inverse image by the canonical projection in $\mathscr{ET}$ is open. When we use the open sets provided by \eqref{23}, the topology in the arrow of time is exactly the same one known from the standard topology of intervals in $\mathds{R}$. Regarding the non-Hausdorff character of the topology here constructed, it is in perfect agreement with the conception of a pre-relativistic space-time as a fiber bundle $\left(\mathds{E}^1\times\mathds{E}^3, \pi, \mathds{E}^1  \right)$, where $\mathds{E}^1$ and $\mathds{E}^3$ may be seen as the one and three dimensional Euclidean spaces, respectively. The map $\pi: \mathds{E}^1\times\mathds{E}^3\rightarrow\mathds{E}^1$ is the canonical projection given by $\pi(t,x)=t$. By the continuity of the projection map, the topology on the total space $\mathds{E}^1\times\mathds{E}^3$ is given by $\mathcal{T}_{\mathds{E}^1\times\mathds{E}^3}=\{I\times\mathds{E}^3, \ I\in\mathcal{T}_{\mathds{E}^1}\}$. Under this description, the worldline of a particle is represented by a cross-section $\gamma: \mathds{E}^1\rightarrow\mathds{E}^1\times\mathds{E}^3$ which is a continuous map such that $\pi\circ\gamma=id_{\mathds{E}^1}$. Although the time interval topology is coarser than the product topology on $\mathds{E}^1\times\mathds{E}^3$, they induce the same topology on the time arrow, that is, that of $\mathds{E}^1$. Furthermore, the sets $F_t:=\pi^{-1}(\{t\})=\{t\}\times\mathds{E}^3\cong\mathds{E}^3, \ t\in\mathds{E}^1$ may be seen as the leaves in the foliation given by $\mathcal{F}:=\{F_t\}_{t\in\mathds{E}^1}$. The description of pre-relativistic space-time with its topological or geometric properties is not our main focus on this paper, however for insightful discussions see \cite{kopczynski_spacetime_1992}. 

\section{The transition from classical to relativistic re-gime and vice versa}

Let us discuss what are the consequences for both the arrow of time and time interval topology when we accept the special relativity postulates. Our first question is the following: is there any fundamental difference in the time structure discussed so far when one inserts the speed of light as a superior bounded and invariant scale? And what about the time interval topology? How it may be accommodated to reflect this new feature? 

The special relativity theory is based upon two known postulates. The first one guarantees a universal character of physics, as its laws must be the same for all observers in any (inertial) frame of reference. In addition, the second postulate graduates the speed of light $c$ as a constant, maximum and invariant scale that all observers should agree on. One of its consequences, of interest to this work, is that the very notion of simultaneity is deeply disrupted from the one we have constructed in here. The technical details can be seen in many different references \cite{ugarov, bernbook}. The central idea we would like to explore is related to a trichotomy so imposed by the postulates \cite{oneil}. Given a pair of distinct events, say $e_1$ and $e_2$, one, and only one of the three options holds true.

\textbf{\textit{i)}} $e_1-e_2$ is time-like. It means that we may prepare a free particle (used as clock and denoted by $\tau(\cdot, \cdot)$) to witness both $e_1$ and $e_2$. We call the corresponding temporal distance $\tau(e_1,e_2)$. 

\textbf{\textit{ii)}} $e_1-e_2$ is light-like. The events are connected by a light signal. Operationally, we could think of $e_1$ and $e_2$ as an emission followed by the reception of the signal.

\begin{figure} [H]
	\centering
	\includegraphics[scale=0.3]{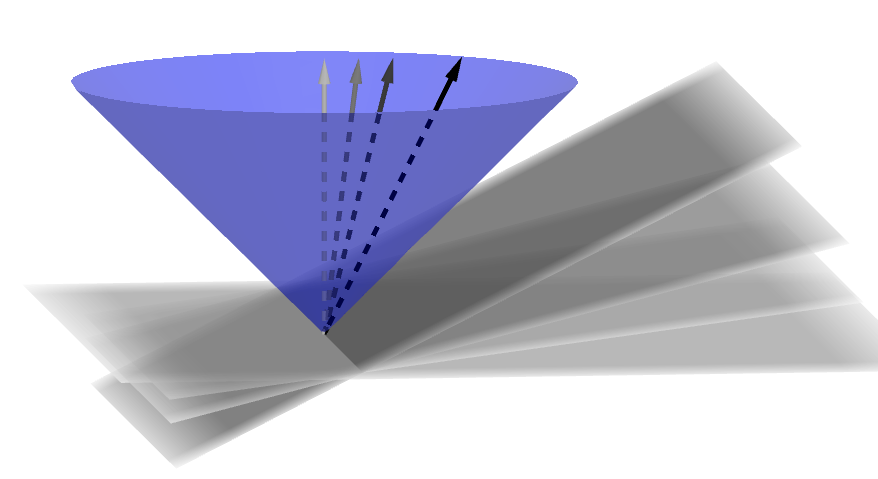}
	\caption{Space-time instants with corresponding arrows of time.}
	\label{arrows}
\end{figure}

\textbf{\textit{iii)}} $e_1-e_2$ is space-like. There is a particular frame of reference in which $e_1$ and $e_2$ happen simultaneously in points, say $E_1$ and $E_2$. Hence we could define the spatial distance of $e_1$ and $e_2$ by $d(E_1, E_2)$. 
Since simultaneity is now relative, events that live in a particular instant may no longer be simultaneous for another observer that moves with respect to the latter. This indicates that the arrow of time is not unique anymore - every observer has its own. In fact, we have an entire family of times $\left ( t_{\vec{V}}\right )_{\vert\vec{V}\vert \in [0,c]}$, parametrized by the relative velocity $\vec{V}$ between frames \cite{rovelli_quantum_gravity}. We point out that there is no contradiction with our previous construction. The only difference lays on its locality, that is, any observer should build its own. Fig. \ref{arrows} shows space-time slices representing instants, each one with its corresponding time line. 

\begin{figure} [H]
	\centering
	\includegraphics[scale=0.2]{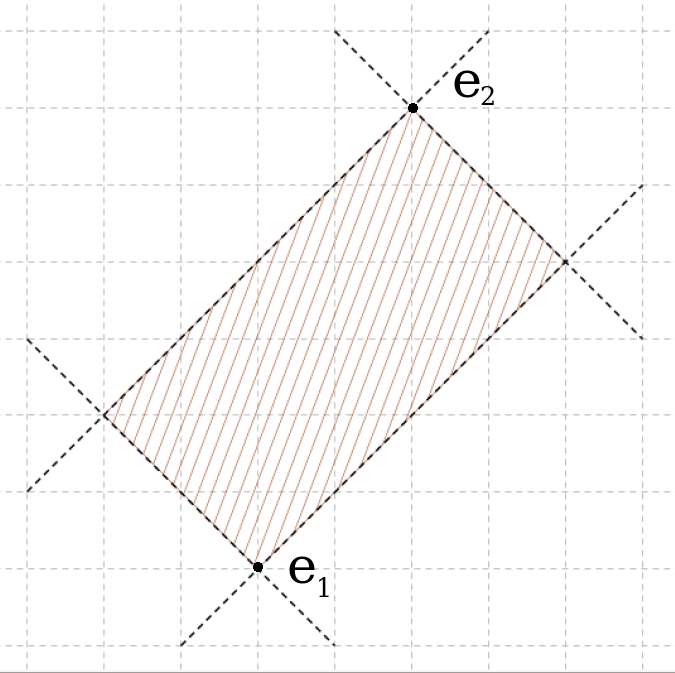}
	\caption{Representation of an open set $\mathcal{A}(e_1,e_2)$ in space-time.}
	\label{openset}
\end{figure}

\begin{figure} [H]
	\centering
	\includegraphics[scale=0.3]{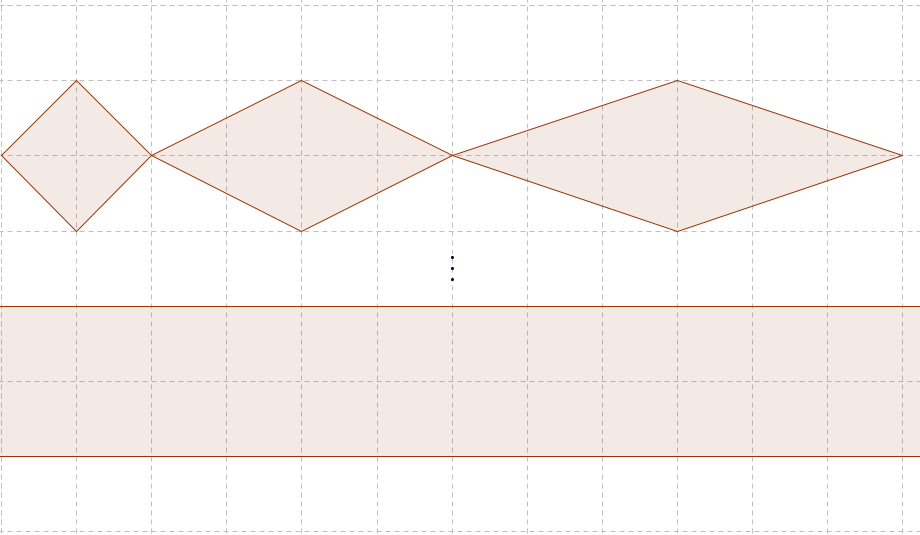}
	\caption{Space-time topology behaviour through $c \rightarrow +\infty$ limit.}
	\label{limiti_topology2}
\end{figure}

We finish this Section with some comments on the topological space-time structure one can obtain from our previous arguments. Due to the trichotomy implied by the special relativity postulates, given an event $e_1$, only a subset of events in the space-time can be compared to it in the sense of past and future. The future $\mathscr{F}(e_1)$ of $e_1$ is defined by the subset $e \in \mathscr{ET}$ such that $\tau(e_1,e)$ is well defined and the clock $\tau(\cdot, \cdot)$ witnesses first $e_1$ and then $e$. Conversely, the past $\mathscr{P}(e_1)$ of $e_1$ is defined by the subset $e \in \mathscr{ET}$ such that $\tau(e, e_1)$ is properly defined and $\tau(\cdot, \cdot)$ witnesses $e$ first. So, let $e_2$ be in the future of $e_1$. Open sets are defined in the same way we have done before,
\begin{equation}
	\mathcal{A}(e_1, e_2):= \mathscr{F}(e_1)\cap \mathscr{P}(e_2),
\end{equation}
and the set 
\begin{equation}
	\mathcal{B} = \{ \mathcal{A} \subset \mathscr{ET}\vert \mathcal{A} \mbox{ is an open set} \}    
\end{equation}
is the corresponding basis for the time interval topology\footnote{This topology is also known in the literature as the ``Alexandrov interval topology'' \cite{alexandrov_1959, sen_1999}.}, which, in this case, is Hausdorff. Operationally, it is impossible to prepare a particle with speed that exceeds $c$, the speed of light. It restrains the structure of open sets, depicted in Figure \ref{openset}.
If greater and greater speeds are allowed for free particles used as clocks, then more spatially distant events can be called time-like. In the limit $c \rightarrow +\infty$, the open sets $\mathcal{A}(e_1, e_2)$ degenerates to $I(e_1, e_2)$, as expressed in the Figure \ref{limiti_topology2}.

\section{Conclusion}

The central objective of this paper was to answer the very question presented already in the title: is the time line the real line? It was motivated by the seminal paper ``Why is the Euclidean Line the Same as the Real Line'' by Prof. R. Sen \cite{sen_1999}. As we have seen, there are good clues that time can be properly described by an oriented (in the sense of past and future) differentiable one-dimensional manifold. To see that, we followed a path twofold degenerated, composed essentially by an operational philosophy, and whenever necessary, we have taken leaps of abstraction, leveraging a fully mathematical overview.

The entire operational sector was devoted to constructing time intervals as a physical quantity. We started defining an experimental procedure to obtain the corresponding values with a clepsydra and then completed it using free particles as clocks. It should be mentioned though, that yet the operational construction of time intervals was accomplished with a clepsydra, we are free to choose any experimental apparatus, as long as it allows for a complete formalized operational description. Due to the rational values any experimental measurement provides, the arrow of time was first identified as the normed vector space $(\mathds{Q}(s), \vert\cdot\vert)$. Then we concluded our main result, $\overline{\mathds{Q}(s)}=\mathds{R}(s)$, invoking the unique completion (up to isometries) every normed vector space admits. Although the completion of $\mathds{Q}(s)$ is a mathematically perfectly reasonable step, its physical necessity is debatable, see \cite{bombelli.1987} for an alternative construction with a discrete space-time. We reinforce that this transition is only a mathematical procedure, detached from what is indeed physically measurable.

Unfortunately, we fall here in a particular void: the arrow of time was constructed first as $\mathds{Q}(s)$ and extended by mathematical completion to $\mathds{R}(s)$. Given two events, say $e_1$ and $e_2$, we will always measure a rational $\delta t(e_1, e_2)$. Since the pair of events has already elapsed, it is no longer possible to find $e$ such that $e_1 \prec e  \prec e_2$. We may reproduce events $e'_1$ and $e'_2$ with the same time lapse as the one defined by $e_1$ and $e_2$, and a third event $e'$ satisfying $e'_1 \prec e'  \prec e'_2$. However, we can never be sure if there is any hole in between $e_1$ and $e_2$ as time keeps flowing. Compared to the static case, as in \cite{sen_1999}, one can always return to a particular segment in order to analyse its ``experimental'' topological structure. Despite this fact, that is, the completion (as a pure mathematical operation) of the arrow of time, the question whether we may characterize it as $\mathds{R}(s)$ is presently unknown. There are other possible number systems that may as well be used to do the job of modeling empirical reality. The interested reader may find a compelling discussion in \cite{matt.visser}.

During the development of our main result discussed above, we have also exposed the space-time topological structure. Figure \ref{arrow} indicates a foliation, under our description of fiber bundles. This fact was lightly introduced at the end of Section \ref{sec5}. In the pre-relativistic context, the time interval topology was closely related to the causal structure. Curiously enough, in this case, the topology is non-Hausdorff. Moreover, it can be noticed that the subspace topology generated on hyperplanes of simultaneous events is the trivial topology, since given a hyperplane $\{t\}\times\mathds{E}^3$, its subspace topology is $U\cap\{t\}\times\mathds{E}^3$, where $U$ is open in the time interval topology. Thus, the topology induced on the hyperplane of simultaneous events is $\mathcal{T}_{induced}=\{\emptyset,\{t\}\times\mathds{E}^3\}$, which is the trivial topology. It may be unwanted from the physical point of view, but that is what we can do with the little physical structure we considered the pre-relativistic space-time to have. By the way, nothing prevents us from considering the Euclidean topology on $\mathds{E}^3$, since it is another topological space that is somehow glued to every time value under the fiber bundle description. We are compromised with physical implications that come solely due to time considerations. For physical considerations concerning space, we must introduce physical concepts to induce a physical topology on this submanifold, which could very well be the Euclidean topology. For more details, see the first chapter on \cite{arnold_mathematical_1989}.

The transition from classical to relativistic regimes was also discussed. Min-kowski space-time viewed as $(\mathds{R}^4, \eta)$ assumes, in general, the Euclidean topology on $\mathds{R}^4$ \cite{naber.relativity.2012}. In this case, our time interval topology on $\mathds{E}^1\times\mathds{E}^3$ is not homeomorphic, hence not diffeomorphic, to $(\mathds{R}^4, \eta)$. However, the time interval topology on the relativistic space-time distorts the infinite open boxes to diamond-like shape open sets, which are the basis for a topology on $\mathds{E}^1\times\mathds{E}^3$, and is equivalent to the product topology on the Euclidean spaces $\mathds{E}^1$ and $\mathds{E}^3$. The connection between open sets in each case can be seen through the limit $c \rightarrow +\infty$: the diamond open sets degenerate to the boxes. This transition was depicted in Figure \ref{limiti_topology2}.

Perhaps we should ``forget time'' \cite{rovelli_forget_2009} or see it as an effective physical quantity - a particular subset (satisfying technical conditions) of an ordered number field equipped with a (quasi) distance function \cite{gallego_torrome_general_2021}. In any case, this paper sheds light in both the mathematical and physical structures it bears, in both classical and relativistic regimes. 

\section*{Acknowledgement}

This work is supported by Programa Institucional de Bolsas de Inicia\c{c}\~ao Cient\'ifica - XXXII BIC/UFJF- 2019/2020, project number ID46462.

BFR would like to express his gratitude to Prof. V. Gueorguiev, who read the first version of this manuscript and gave many useful suggestions to improve its content.

\section*{Data Availability}

Data sharing not applicable to this article as no datasets were generated or analyzed during the current study.

\end{document}